# Spectral density and calculation of free energy

**Leonid Litinskii and Boris Kryzhanovsky**

Scientific Research Institute for System Analysis RAS,
Center of Optical Neural Technologies
Nakhimov ave, 36-1, Moscow, 117218, Russia
litin@mail.ru , kryzhanov@mail.ru

For planar and cubic Ising models, we examined two ways of approximation of a spectral density that describes a degeneracy of energy levels. We approximated the exponent of the spectral density by polynomials of even degrees and using our n-vicinity method [8, 9]. According our analysis, the free energy is almost independent of the chosen method of approximation. However, its derivatives depend on the way of approximation and substantially differ in the neighborhood of a critical temperature. Our calculations showed that when approximating by polynomials the system necessarily finds itself in the ground state at a finite temperature, which is forbidden. The n-vicinity method approximates the derivatives of the free energy correctly for the cubic Ising model and it works poorly in the planar case.

## 1. Introduction

The most natural way to calculate a partition function $Z_N(\beta) = \sum_{\text{all } \mathbf{s}} \exp[-\beta N E(\mathbf{s})]$, $\beta = 1/T$, of a system consisting of a large number of spins $N \gg 1$ is as follows[1]. Let $\{E_r\}$ be the set of all the different energies of the states $\mathbf{s} = (s_1, s_2, ..., s_N)$ and $D_r$ is the degeneracy order of the energy level $E_r$ that is the number of the states with such an energy. The same as in physics the spectral density is the set $\{D_r\}$, and $\sum_r D_r = 2^N$.

When we know $\{E_r\}$ and $\{D_r\}$, we write the partition function $Z_N(\beta)$ as a sum over all the different energies

$$Z_N(\beta) = \sum_r D_r \exp(-\beta N E_r). \tag{1.1}$$

However, we know the spectral density $\{D_r\}$ only for some special cases. Namely, they are

- the mean field model where $q$ nearest neighbors are taking into account [1]:

$$E_r = -\frac{qJ}{2} \frac{(N-2r)^2 - N}{N(N-1)}, \quad D_r = 2C_N^r, \quad r = 0,1,...N/2;$$

- the one-dimensional Ising model [2]:

$$E_r = -J\left(1 - \frac{4r}{N}\right), \quad D_r = 2C_N^{2r}, \quad r = 0,1,...N/2.$$

In the both formulas, $J$ is the interaction constant between neighboring spins.

---

[1] The factor N in the exponent means that we calculate the energy $E(\mathbf{s})$ per one spin (see Eq. (1.4)).

We can try to approximate the spectral density $\{D_r\}$ with the aid of one or another method. In the present paper, we compare the results obtained when approximating $\{D_r\}$ by an approximation by polynomial and the n-vicinity method.

In Sections 2 and 3, we discuss the planar Ising model. We begin Section 2 with calculation of the free energy using the exact Onsager result [3]. After that, basing on the obtained free energy we restore the exact spectral density [4-6]. We used it as a reference when comparing two methods of approximation analyzed in this publication.

Concluding Section 2, following [7] we discuss an approximation of the spectral density by polynomials. The main points of the approximation by polynomials are as follows. Let us agree to write the spectral density as an exponent $D_r = \exp(N\varphi_r)$, where the quantity $\varphi_r$ in the exponent we call the exponent of the spectral density. In [7] for the planar and cubic Ising models, they simulated the distributions $\{D_r\}$ and found that for large dimensions ($N \sim 10^3$) they were bell-shaped curves symmetrical around the point $E = 0$. They had long "tails" pressed to the abscissa axis. Comparing $\varphi_r$ with a Gaussian exponent $g_r = \ln 2 - E_r^2/2\sigma^2$ ($\sigma^2$ is a variance), they found that at the tails the value of $\varphi_r$ differed from $g_r$ significantly. In other words, the Gaussian density does not approximate the simulated distribution correctly. After that, in [7] they tried to approximate $\varphi_r$ most accurately by polynomials of even degrees:

$$\varphi_r = \text{const} + a \cdot E_r^2 + b \cdot E_r^4 + c \cdot E_r^6. \tag{1.2}$$

In the case of the planar Ising model, for a correct approximation it is enough to use the first three summons in Eq. (1.2), however for the cubic Ising model all the summons have to be taken into account. In [7], fitting the constants in Eq. (1.2) the authors obtained a suitable approximation of the spectral density, calculated the free energy and analyzed its behavior. The critical value of the inverse temperature they found within the accuracy of 12%. We used the approximation (1.2) and repeated computer simulations presented in [7]. After a detailed analysis, we determined that at a finite temperature the system found itself in the ground state (and that contradicts to the general physical principles). In addition, the approximation (1.2) leads to a phase transition of the first kind, which contradicts to the exact Onsager solution. We conclude Section 2 by a presentation of the obtained results.

In Section 3, we present another way of approximation of $\{D_r\}$. This is the n-vicinity method we developed in [8, 9]. In the framework of this method, we choose an initial configuration $\mathbf{s}_0 \in \mathbf{R}^N$ and all other configurations we group in its n-vicinities $\Omega_n$ each of which contains all configurations that differ from $\mathbf{s}_0$ in opposite values of $n$ spins: $\Omega_n = \{\mathbf{s} : (\mathbf{s}, \mathbf{s}_0) = N - 2n\}$, $0 \le n \le N$. Let $D^{(n)}(E)$ be a true distribution of energies of the states belonging to the n-vicinity $\Omega_n$. In the general case, we do not know $D^{(n)}(E)$. However, we can calculate accurately its mean $E_n$ and variance $\sigma_n^2$ in terms of the parameters of the connection matrix and the configuration $\mathbf{s}_0$ [8]. According the n-vicinity method we replace an unknown exact distribution $D^{(n)}(E)$ by the Gaussian distribution

$$D^{(n)}(E) \sim \exp\left(-\frac{N}{2}\left(\frac{E - E_n}{\sigma_n}\right)^2\right). \tag{1.3}$$

Calculating the weighted sum of the partial densities (1.3), we approximated the spectral density by

$$D_{nv}(E) \sim \sum_{n=0}^{N} C_N^n \cdot \exp\left(-\frac{N}{2}\left(\frac{E-E_n}{\sigma_n}\right)^2\right).$$

Here the subscript "nv" denotes the n-vicinity. Justification of the n-vicinity method we published in [8], the boundaries of its applicability for Ising models on hypercubes we analyzed in [9]. In Section 3, for the planar Ising model we discuss the results following from this approach. It also predicts the phase transition of the first kind but there is no a transition to the ground state at a finite temperature.

In Section 4, we present the results relating to the cubic Ising model. Because of the absence of an exact expression for the free energy, we can only compare the both approximate spectral densities $D_{nv}(E)$ and $D_{pl}(E)$ and the corresponding free energies (the subscript "pl" denotes the approximation by polynomials). As for the planar Ising model in this case the approximation by polynomials also provides incorrect results. On the contrary, the approximation by means of the n-vicinity method coincides well with the widely accepted results of computer simulations [10].

The discussion and conclusions are in Section 5. We explain why the transition to the ground state at a finite temperature is an unavoidable defect intrinsic to the approximation by polynomials. We also discuss a possibility to improve the n-vicinity method.

In Appendix we present the details of the calculations.

**Notations.** In what follows we calculate the energy per one spin, i. e.

$$E(\mathbf{s}) = -\frac{1}{2N}\sum_{i\neq j}^{N} T_{ij} s_i s_j . \tag{1.4}$$

We analyze the simplest variants of the planar and the cubic Ising models. This means that only the nearest spins interact, the interaction constant $J=1$, and the boundary conditions are periodic. In the n-vicinity method we take the ground state as an initial configuration: $\mathbf{s}_0 = (1,1,...,1) \in R^N$. Each spin has $q = 2d$ nearest neighbors; $d$ is a dimension of the lattice. The energy of the ground state is:

$$E_0 = E(\mathbf{s}_0) = -\frac{q}{2}. \tag{1.5}$$

In the case of the planar Ising model $q=4$ and $E_0 = -2$; for the cubic Ising model $q=6$ and $E_0 = -3$. The spectral density is defined on the interval $[E_0, |E_0|]$. The exact free energy we denote by $f(\beta)$ and the free energies corresponding to approximations by polynomials and by the n-vicinity method we denote by $f_{pl}(\beta)$ and $f_{nv}(\beta)$, respectively.

## 2. Planar Ising model: approximation by polynomials

In this Section, we, first, calculate the exact spectral density

$$D(E) = \exp(N \cdot \Psi(E)) \tag{2.1}$$

and, second, approximate the function $\Psi(E)$ by a polynomial of the forth degree $\Psi_{pl}(E)$. Comparing the function $\Psi(E)$ and $\Psi_{pl}(E)$ as well as the graphs for $f(\beta)$ and $f_{pl}(\beta)$, we can define advantages and disadvantages of the polynomial approximation.

*2.1. Exact spectral density: general formulas*

The standard expression for the dimensionless free energy $f(\beta)$ in terms of the partition function $Z_N(\beta)$ has the form

$$f(\beta) = -\lim_{N \to \infty} \frac{\ln Z_N(\beta)}{N}. \qquad (2.2)$$

Here $\beta = 1/T \in (0, \infty)$ is the inverse temperature. Suppose we know the free energy. For example, analyzing the planar Ising model we can use the exact Onsager result [3]. Then we can calculate the exponent of the spectral density according the formulas [4–6]:

$$\Psi(E) = \beta E - f(\beta), \text{ where } E = f'(\beta). \qquad (2.3)$$

From Eqs. (2.3), it follows that

$$\Psi'(E) = \beta, \text{ and } \Psi''(E) = \left(f''(\beta)\right)^{-1} \leq 0. \qquad (2.4)$$

Note, up to a constant $\Psi(E)$ is the entropy. If we know a free energy of a system Eqs. (2.3) allow us to calculate the exponent of the spectral density and the equations (2.4) to find its derivatives with high accuracy.

Let us list the general properties of the free energy and the spectral density. The free energy is a nonincreasing function. Then equations (2.3) define $\Psi(E)$ for negative energies only. Since in the Ising model $\Psi(E) = \Psi(-E)$, we can continue the function $\Psi(E)$ along the positive semi-axis. In what follows we examine the energies from the interval $E_0 \leq E \leq |E_0|$. Next, when $\beta \to \infty$ the spin system approaches the ground state. The energy of the ground state we denoted as $E_0$ (see Eq. (1.5)) and consequently $\lim_{\beta \to \infty} E(\beta) = E_0$. In other words, when $\beta \to \infty$ the free energy $f(\beta) \to \beta E_0$ and from Eq. (2.3) it follows that

$$\lim_{\beta \to \infty} \left(\beta E - f(\beta)\right) = \lim_{E \to E_0} \Psi(E) = 0.$$

In other words, at the ends of the energy interval the exponent $\Psi(E)$ must tend to zero.

As known, the second derivative of the free energy is proportional to the variance of the energy distribution at the given temperature taken with the opposite sign. The variance cannot be negative and this defines the sign in the inequality (2.4). Note, since the second derivative $\Psi''(E)$ is not positive the first derivative $\Psi'(E)$ must be a monotonically nonincreasing function.

In Fig. 1, we show the graphs of the exponent $\Psi(E)$ and its two derivatives for the planar Ising model. The graphs demonstrate the properties of these functions listed above. The author of paper [6] showed that at the left end of the interval near $E_0 = -2$ the function $\Psi(E) \sim -(E - E_0)\ln(E - E_0)$. Then its derivatives are $\Psi'(E) \sim -\ln(E - E_0)$ and

$\Psi''(E) \sim -1/(E - E_0)$. In other words, at the left end of the interval, the first derivative $\Psi'(E)$ logarithmically tends to the positive infinity and the second derivative tends to the negative infinity.

The first derivative $\Psi'(E)$ is a smooth function that has an inflection point at $E_c = E_0/\sqrt{2}$. In what follows we describe only the left sides of all the figures where we show the behavior of the spectral density. It is significant that at that point the second derivative is equal to zero: $\Psi''(E_c) = 0$. Taking into account the inequality (2.4) we obtain that at the corresponding value $\beta = \beta_c$ the second derivative of the free energy has an infinite discontinuity: $f''(\beta) \to \infty$ when $E \to E_c$. This means that at the point $\beta = \beta_c$ a phase transition of the second kind takes place. Near the energies $\pm E_c$ the exponent $\Psi''(E)$ becomes equal to zero very rapidly. In Fig. 1, the singular regions resemble tiny horns. Near the origin of coordinates $E = 0$ the asymptotics of the function $\Psi(E)$ is (see [6])

$$\Psi(E) \sim \ln 2 - E^2/4 \implies \Psi'(E) = -E/2, \ \Psi''(E) = -1/2. \tag{2.5}$$

These expressions will be useful in what follows.

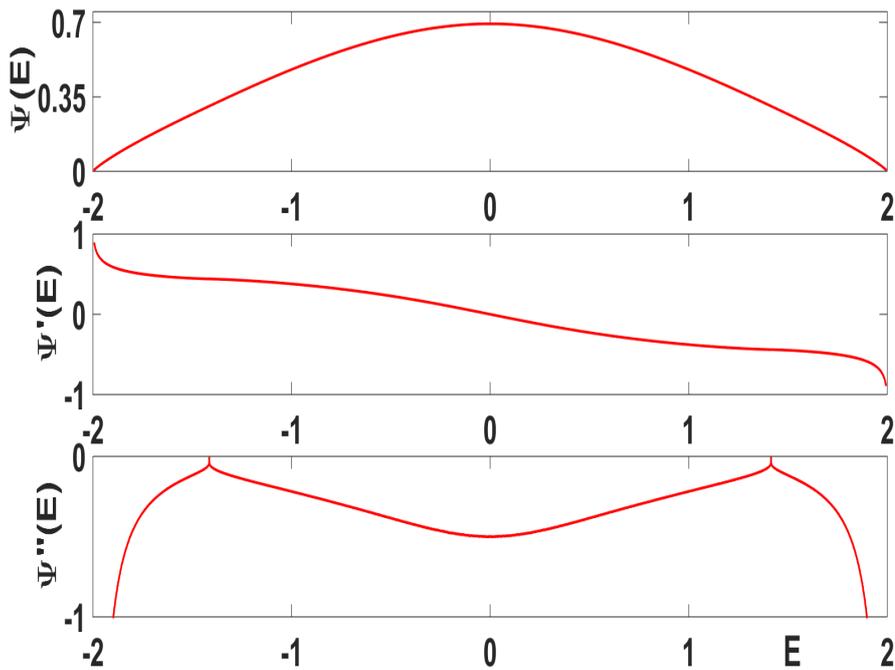

Fig. 1. Exponent $\Psi(E)$ of spectral density $D(E)$ and its derivatives for planar Ising model.

*2.2. Approximation of spectral density by polynomials*

Simulating the spectral density for the planar Ising model authors of [7] showed that with high accuracy they could approximate it by the fourth order polynomial

$$D_{pl}(E) \sim \exp(N \cdot \Psi_{pl}(E)), \ \Psi_{pl}(E) = \ln 2 - a \cdot E^2 + b \cdot E^4,$$
$$a = 1/(2.16001)^2, \ b = 1/(2.09255)^6, \ E \in [-2, +2]. \tag{2.6}$$

In Fig. 2, the left column presents the graphs of the exponents $\Psi(E)$, $\Psi_{pl}(E)$ and their derivatives. In the upper panel, we see the graphs of $\Psi(E)$ and $\Psi_{pl}(E)$, the middle panel shows the derivatives $\Psi'(E)$ and $\Psi'_{pl}(E)$, and in the bottom panel there are the second derivatives $\Psi''(E)$ and $\Psi''_{pl}(E)$. In this figure, solid lines correspond to the exact exponent $\Psi(E)$ and dash-dotted lines to the polynomial $\Psi_{pl}(E)$.

We see that as a whole $\Psi_{pl}(E)$ approximates the exponent $\Psi(E)$ correctly. The curves almost coincide on the interval $-1.5 \le E \le 0$ and they differ near the energy of the ground state $E_0 = -2$.

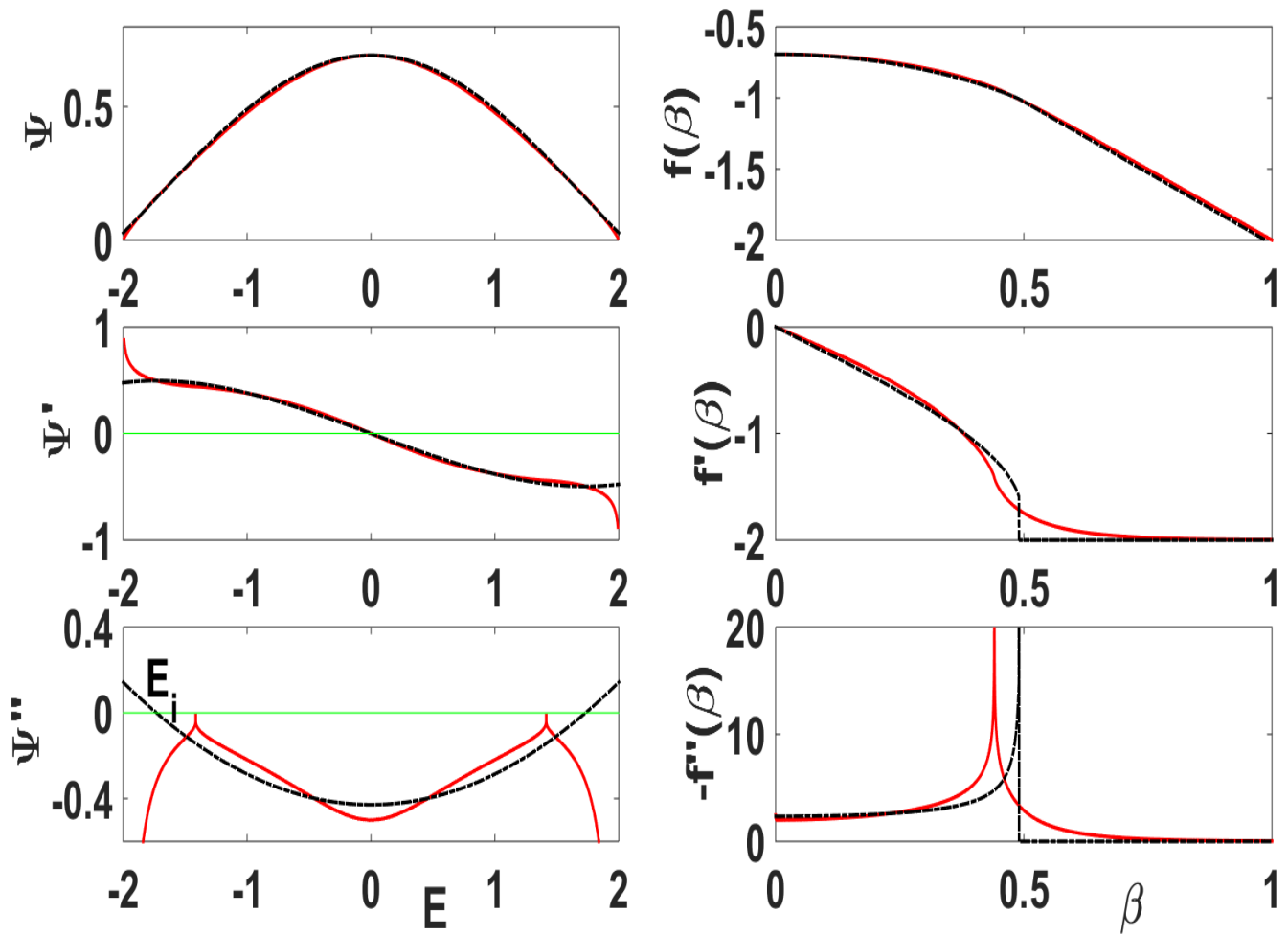

Fig. 2. Comparison of exact solution and approximation by polynomials for planar Ising model. In left column are exact exponent $\Psi(E)$, its approximation $\Psi_{pl}(E)$ and their derivatives (solid lines and dash-dotted lines, respectively). In right column are free energies $f(\beta)$, $f_{pl}(\beta)$ and their derivatives (solid lines and dash-dotted lines, respectively).

The situation is not so optimistic when we compare the derivatives of these functions (see graphs for $\Psi'(E)$ and $\Psi'_{pl}(E)$ in the middle panel of the figure). In general, at the interval $(-1.8, +1.8)$ the graphs are rather close. However, when $E \to E_0$ the first derivative $\Psi'_{pl}(E)$ is finite but $\Psi'(E)$ tends to infinity (see Appendix). We will see later that the finite value of the derivative $\Psi'_{pl}(E_0)$ is the reason why the system finds itself in the ground state at a finite temperature. Such a behavior of the system makes no physical sense. In the case of the exact spectral density $\Psi'(E) \to \infty$ when $E \to E_0$. That is the reason why the system only tends to the ground state asymptotically when $\beta \to \infty$ and it does not reach it at any finite value of $\beta$.

The left bottom panel of Fig. 2 shows the graphs of the second derivatives $\Psi''_{pl}(E)$ and $\Psi''(E)$. The parabolic curve $\Psi''_{pl}(E)$ describes badly the complicated behavior of $\Psi''(E)$. At the point

$$E_i = -\sqrt{\frac{a}{6b}} \approx -1.7318 \qquad (2.7)$$

the second derivative $\Psi''_{pl}(E)$ is equal to zero: $E_i$ is the point of inflection of the curve $\Psi_{pl}(E)$. To the left of the energy $E_i$ the second derivative $\Psi''_{pl}(E)$ is positive. Since according the inequality (2.4), the second derivative of the spectral density cannot be positive, the interval $[E_0, E_i]$ is forbidden. This means that only in the inner interval $E_i \leq E \leq |E_i|$ we can approximate the exponent $\Psi_{pl}(E)$ by the polynomial $\Psi_{pl}(E)$.

## 2.3. Equation of state and its solution

Let us show how the properties of the spectral density define the behavior of the free energy. Now in place of $\Psi_{pl}(E)$ we use $f_0(E) = -\Psi_{pl}(E)$:

$$f_0(E) = -\ln 2 + a \cdot E^2 - b \cdot E^4. \qquad (2.8)$$

Then the partition function has the form

$$Z_{pl}(\beta) \sim \int e^{-\beta NE} D_{pl}(E) dE = \int e^{-N \cdot [f_0(E) + \beta E]} dE \sim e^{-N \cdot f(\beta)}.$$

Here $f(\beta)$ is the dimensionless free energy (2.2). For any given $\beta$ we calculate $f(\beta)$ minimizing the function $f_0(E) + \beta \cdot E$:

$$f(\beta) = \min_E \left( f_0(E) + \beta \cdot E \right).$$

The equation of state allowing us to determine the value of the free energy has the form

$$f_0'(E) + \beta = 0. \qquad (2.9)$$

Now we see an evident shortcoming of the approximation by polynomials: the derivative $f_0'(E)$ is bounded over all the interval $E_0 \leq E \leq 0$ and the equation of the state (2.9) has no solutions when $\beta > -\max f_0'$.

Let us analyze graphically the behavior of the solution of equation (2.9). On the upper panels of Figs. 3a) and 3b) we show the graphs of the functions $f_0(E)$ and $f_0(E) + \beta \cdot E$ for two different values of the parameter $\beta$: $\beta = 0.1$ and $\beta = 0.4$. On the bottom panels are the derivatives $f_0'(E)$ and $f_0'(E) + \beta$. The graph of the functions $f_0'(E) + \beta$ is simply the graph of the derivative $f_0'(E)$ shifted upwards by the quantity $\beta$.

According Eq. (2.8) the minimum of the function $f_0(E)$ is at the point $E_z = 0$, and at that $f_0(0) = -\ln 2$. This function has a maximum at the point $E_M = -\sqrt{a/2b} \approx -3$ that is outside the energy interval $[-2, 0]$; this point will be useful in what follows. In Eq. (2.7) we defined the point of inflection $E_i$ of the function $f_0(E)$. It is important that the point of inflection of the function $f_0(E) + \beta \cdot E$ does not depend on $\beta$ and it remains fixed when $\beta$ changes.

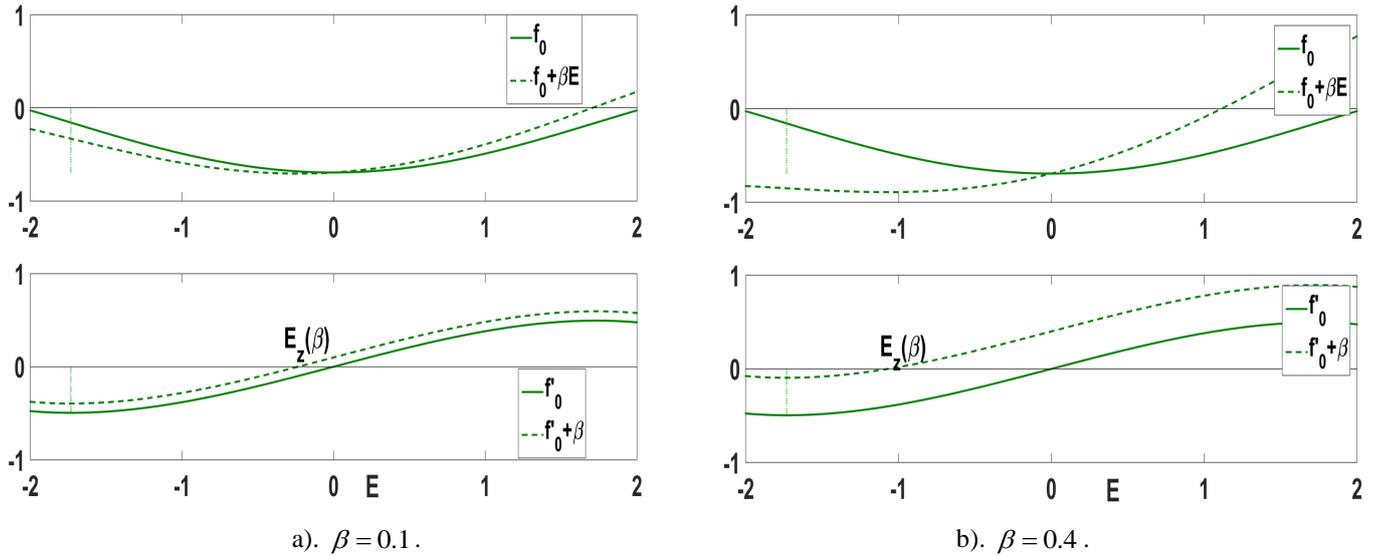

a). $\beta = 0.1$.   b). $\beta = 0.4$.

**Fig. 3.** Functions $f_0(E)$ and $f_0(E) + \beta \cdot E$ (upper panels), and derivatives $f_0'(E)$ and $f'_0(E) + \beta$ (bottom panels) for two values of parameter $\beta$. Vertical dotted straight line marks point of inflection $E_i$ (2.7).

Let us follow the minimum of the function $f_0(E) + \beta \cdot E$. At first, when $\beta = 0$ the minimum is at the point $E_z = 0$. When adding the term $\beta \cdot E$ to $f_0(E)$ we obtain an asymmetrical function (the dashed lines on upper panels in Fig. 3). Then the minimum point $E_z(\beta)$, shifts from zero to the left. The graphs of the derivatives $f_0'(E) + \beta$ show that when $\beta$ increases this point always shifts from zero to $E_0$ (see the bottom panels in the figure). On the contrary, the point of the maximum $E_M(\beta)$ shifts from the left to the right toward the origin of coordinates. When $\beta = 4E_i a/3 \approx 0.4949$ the points $E_z(\beta)$ and $E_M(\beta)$ have to coincide at the point of inflection $E_i$. However that does not occur because when $\beta = \beta_j = 0.49032$ the left end of the graph of the function $f_0(E) + \beta \cdot E$ shifts down so far that the value of this function becomes equal to its value at the minimum point $E_z(\beta)$:

$$f_0(E_0) + \beta_j \cdot E_0 = f_0\left(E_z(\beta_j)\right) + \beta_j \cdot E_z(\beta_j). \tag{2.10}$$

In Fig. 4 on the upper panel, we show the behavior of the function $f_0(E) + \beta_j \cdot E$ near the left end of the interval on an enlarged scale. When $\beta = \beta_j$ the global minimum of the function "jumps" to the boundary of the interval

$E_0 = -2$ (the subscript "j" denotes "jump"). After that, the global minimum of the function $f_0(E) + \beta \cdot E$ remains at the point $E_0$ no matter how much the value of $\beta$ increases.

Such behavior is a result of the finite value of the derivate $f_0'(E_0)$ at the left end of the interval. When $\beta$ is sufficiently large the value of the derivative here becomes positive: $f_0'(E_0) + \beta > 0$. At that, the minimum of the function $f(\beta) = f_0(E) + \beta \cdot E$ is at the point $E_0$ and the system finds itself in the ground state at a finite temperature. However, the same as in the case of the exact solution when $f_0'(E \to E_0) \to -\infty$ there is no finite $\beta$ for which the derivative $f_0'(E_0) + \beta$ is positive.

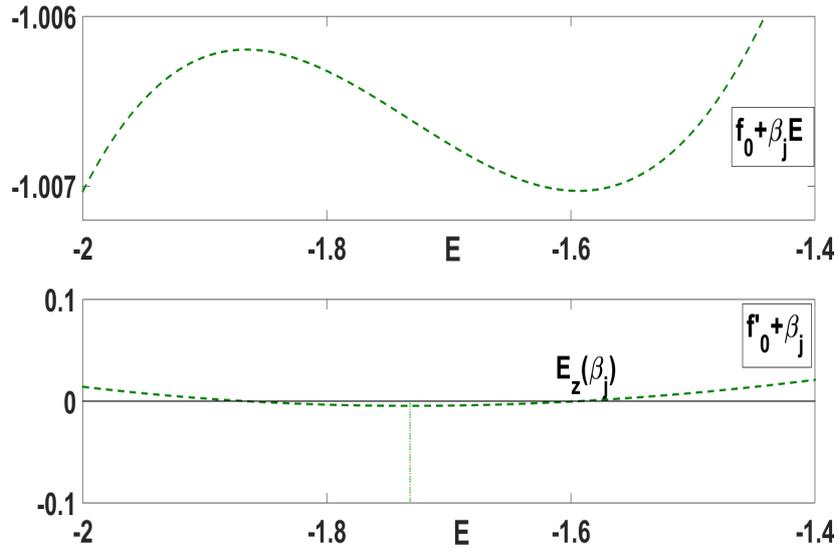

**Fig. 4.** Function $f_0(E) + \beta \cdot E$ (upper panel) and its derivative (bottom panel) for $\beta = \beta_j = 0.49032$. Vertical dotted line marks position of point of inflection $E_i$ (2.7).

*2.4. Free energy*

In the right panels of Fig. 2, we showed the graphs of the free energies $f(\beta)$ and $f_{pl}(\beta)$ as well as their derivatives with respect to $\beta$ calculated using the exact spectral density $\Psi(E)$ (solid lines) and its approximation by the polynomials $\Psi_{pl}(E)$ (dash-dotted lines), respectively.

In the upper right panel, we see that for all $\beta$ the graphs of the functions $f(\beta)$ and $f_{pl}(\beta)$ are nearly indistinguishable. This is not true for their derivatives. As it known, the first derivative $f'(\beta)$, being the internal energy $U(\beta)$, is equal to the averaged energy at a given temperature. It is evident that when $\beta \to \infty$ the first derivative $f'(\beta)$ must tend to the energy of the ground state $E_0 = -2$. The solid curve in the middle right panel of Fig. 2 behaves exactly this way: it decreases smoothly from zero and has the point of inflection at $\beta_c \approx 0.4407$. At first the derivative $U_{pl}(\beta) = f'_{pl}(\beta)$ behaves very similarly, but when $\beta = \beta_j \approx 0.49$ it jumps to the minimal value equal to $E_0 = -2$. This jump follows from Eq. (2.10). Under the further increase of $\beta$, the value of $U_{pl}(\beta)$ remains equal to $E_0$. Such behavior of $U_{pl}(\beta)$ contradicts general physical principles. Consequently, when approximating the internal energy $U(\beta)$ we can use $U_{pl}(\beta)$ only when $\beta < 0.4$. The same is true for the second derivatives $f''(\beta)$ and $f''_{pl}(\beta)$, which behave similarly when $\beta$ is less than $\beta_j \approx 0.49$. For larger $\beta$ it is simply meaningless to compare these functions.

## 3. Planar Ising model: n-vicinity method

In this Section, we derive general expressions allowing us to use the n-vicinity method when approximating the spectral density for an Ising model with $q$ nearest neighbors:

$$D_{nv}(E) = \exp(N \cdot \Psi_{nv}(E)).$$

After that for the planar Ising model ($q = 4$) we compare $\Psi_{nv}(E)$ with $\Psi(E)$ obtained in Section 2 and analyze graphs for the free energies and their derivatives. The case $q = 6$ we discuss in Section 4 when analyzing the cubic Ising model.

### 3.1. Approximation of spectral density by n-vicinity method

In Introduction, we explained the main idea of the n-vicinity method [8, 9]. Below we list the expressions allowing us to use it when approximating the spectral density. We choose the ground state as an initial configuration $\mathbf{s}_0 = (1,1,...,1) \in R^N$. All the other configurations are distributed over its n-vicinities $\Omega_n$ (see Introduction). An unknown exact distribution of energies from $\Omega_n$ we replace by a normal distribution with known first two moments $E_n$ and $\sigma_n^2$. Then the partition function takes the form

$$Z_N = \sum_{\mathbf{s}} e^{-\beta N E(\mathbf{s})} = \sum_{n=0}^{N} C_N^n \left( \frac{1}{C_N^n} \sum_{\mathbf{s} \in \Omega_n} e^{-\beta N E(\mathbf{s})} \right). \tag{3.1}$$

Passing to the limit $N \to \infty$ we turn the sum to an integral. For this purpose, we introduce a variable $x = n/N$ ($x \in [0,1]$) and use Stirling's formula $C_N^{xN} \sim \exp(-N \cdot S(x))$, where $S(x) = x \ln x + (1-x) \ln(1-x)$. In Eq. (3.1), we replace the summation over $n$ by integration over $x$ and the averaging over the states from $\Omega_n$ we replace by integration over the normal density with the mean $E_x = \lim_{N \to \infty} E_n$ and the variance $\sigma_x^2 = \lim_{N \to \infty} \sigma_n^2$:

$$\frac{1}{C_N^n} \sum_{\mathbf{s} \in \Omega_n} e^{-\beta N E(\mathbf{s})} \to \frac{1}{\sqrt{2\pi}\sigma_x} \int_{E_0}^{|E_0|} \exp\left[-\beta NE - \frac{1}{2}\left(\frac{E - E_x}{\sigma_x}\right)^2\right] dE.$$

We use Eq. (1.4) to calculate the energy $E(\mathbf{s})$; the energy $E_0 = E(\mathbf{s}_0)$ we defined in Eq. (1.5). Then

$$E_x = E_0(1-2x)^2, \quad \sigma_x^2 = 8qx^2(1-x)^2 / N.$$

Now we transform Eq. (3.1) to the form

$$Z_N(\beta) \sim \int_{E_0}^{|E_0|} e^{-\beta NE} dE \int_0^1 C_N^{xN} \exp\left(-\frac{1}{2}\left(\frac{E - E_x}{\sigma_x}\right)^2\right) dx,$$

and integrating over $x$ obtain the following approximation of the spectral density:

$$D_{nv}(E) \approx \int_0^1 \exp\left(-NS(x) - \frac{1}{2}\left(\frac{E - E_x}{\sigma_x}\right)^2\right)dx = \int_0^1 \exp(-N \cdot \varphi(E,x))dx = \exp(-N \cdot \varphi_m(E)), \quad (3.2)$$

where

$$\varphi_m(E) = \min_{0 \leq x \leq 1} \varphi(E,x) \quad \text{and} \quad \varphi(E,x) = S(x) + \frac{q}{4}\left(1 + \frac{E/E_0 - 1}{4x(1-x)}\right)^2. \quad (3.3)$$

In what follows, we show that as long as the parameter $\varepsilon = E/E_0$ is less than a critical value $\varepsilon_c$ the minimum of the function $\varphi(E,x)$ is at the point $x_0 = 1/2$ and for the energies from the interval $[\varepsilon_c E_0, 0]$ we have

$$\varphi_m(E) = -\ln 2 + \frac{E^2}{q}, \quad \varepsilon_c E_0 < E < 0. \quad (3.4)$$

When $E < \varepsilon_c E_0$ we can find the minimal value $\varphi_m(E)$ of the function $\varphi(E,x)$ numerically. Determining $\varphi_m(E)$ for each $E$, we obtain the required approximation of the spectral density $\Psi(E)$:

$$\Psi_{nv}(E) = -\varphi_m(E). \quad (3.5)$$

*3.2. Planar Ising model*

Let us examine the planar Ising model ($q = 4$) when the parameter $E \in [E_0, 0]$. It is easy to see that when the energy is not very large, that is $E/E_0 < 0.43$, the function $\varphi(E,x)$ has one minimum at the point $x_0 = 1/2$. According to Eq. (3.4) for these energies we have

$$\varphi_m(E) = -\ln 2 + \frac{E^2}{4}. \quad (3.6)$$

Note, when $0.43 E_0 < E < 0$ the expression for $\Psi_{nv}(E)$ coincides with the analytical expression for the spectral density $\Psi(E)$ near $E = 0$ (see Eq. (2.5)).

When $E \approx 0.4325 \cdot E_0$ one extra minimum of the function $\varphi(E,x)$ appears and when $E \approx 0.438 \cdot E_0$ the depth of the new minimum becomes equal to $\varphi(E, x_0)$. At that the coordinate of the global minimum jumps from the point $x_0 = 1/2$ to the new minimum point $x \approx 0.3$. The jump leads to a jump-like change of the derivative of the function $\varphi_m(E)$. When after that the energy $E$ decreases up to $E_0$ the coordinate of the minimum of the function $\varphi(E,x)$ shifts to zero. It is necessary to stress that the depth of the minimum of $\varphi_m(E)$ tends to zero logarithmically (see Appendix).

In the upper panel of the left column of Fig. 5, we show the graphs of the exponent $\Psi(E)$ defined by Eq. (2.3) and its approximation by the function $\Psi_{nv}(E)$, the derivatives of these functions $\Psi'(E)$ and $\Psi'_{nv}(E)$ are on the middle panel, and on the bottom panel are their second derivatives $\Psi''(E)$ and $\Psi''_{nv}(E)$. As previously, solid lines correspond to the function $\Psi(E)$ and its derivatives and dashed lines to the function $\Psi_{nv}(E)$ and its derivatives, respectively.

The same as in Fig. 2, the graphs on the upper panel are very similar. There is a slight difference between $\Psi(E)$ and $\Psi_{nv}(E)$ only when $-1.5 \leq E \leq -0.5$. Outside this interval, the functions are almost identical.

In the middle panel, we show the derivatives $\Psi'(E)$ and $\Psi'_{nv}(E)$. We see that these graphs are also rather close but the function $\Psi'_{nv}(E)$ has a singularity near $E_j \approx 0.438 E_0 = -0.876$. When $E$ decreases from zero to $E_j$, the graph of the function $\Psi'_{nv}(E)$ is a straight line (see Eqs. (3.5) and (3.6)). When $E$ becomes equal to $E_j$ the derivative $\Psi'_{nv}(E)$ jumps a little: its value changes from $\Psi_{nv}{}'(E_{j+}) \approx 0.44$ to $\Psi_{nv}{}'(E_{j-}) \approx 0.4$. After that, when $E$ varies from $E_j$ to $-1.05$ the derivative $\Psi'_{nv}(E)$ decreases. Consequently, the second derivative $\Psi''_{nv}(E)$ at this interval is positive and that is forbidden by the second equation (2.4). In other words, in the framework of the n-vicinity method the energy interval $[-1.05, -0.876]$ is forbidden. When $E < -1.05$ the second derivative $\Psi''_{nv}(E)$ is negative and the n-vicinity is applicable again. When $E$ decreases from $-1.05$ to the boundary value $E_0$ the derivative $\Psi'_{nv}(E)$ changes smoothly and when $E \to E_0$ the derivative $\Psi'_{nv}(E)$ tends to the infinity. This is the reason why the system does not find itself in the ground state at a finite temperature. It only approaches it asymptotically (see the discussion above).

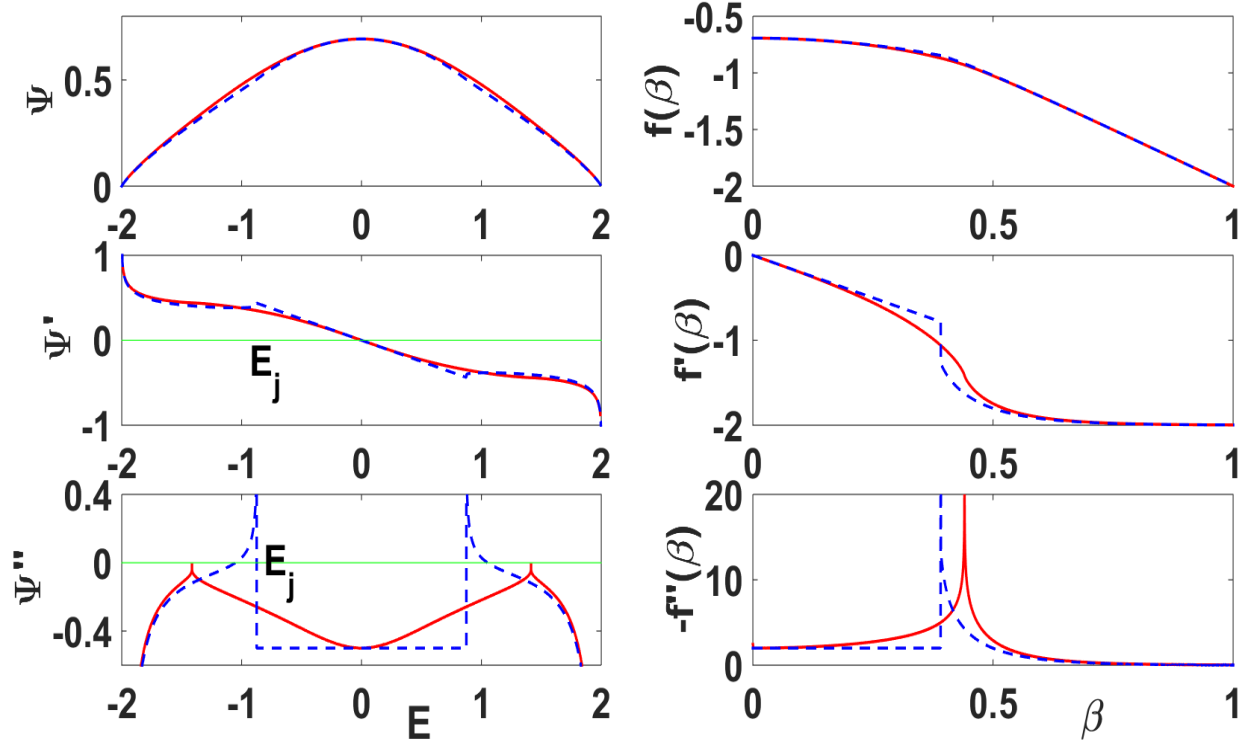

Fig. 5. In left column are $\Psi(E)$ and its derivatives (solid lines) and $\Psi_{nv}(E)$ and its derivatives (dashed lines). In right column are the free energy $f(\beta)$ and its derivatives (solid lines), and $f_{nv}(\beta)$ and its derivatives (dashed line).

In the bottom panel of Fig. 5, we present the graphs of the second derivatives $\Psi''(E)$ and $\Psi''_{nv}(E)$. We clearly see the region of forbidden energies $[-1.05, -0.876]$ where the second derivative $\Psi''_{nv}(E)$ is positive. When energies are less than $-1.5$ the graph of $\Psi''_{nv}(E)$ approximates correctly the function $\Psi''(E)$ and when $E \to E_0$ the function $\Psi''_{nv}(E)$ tends to $\Psi''(E)$.

Note, when $E_j < E < 0$ the function $\Psi''_{nv}(E)$ is the constant: $\Psi''_{nv}(E) \equiv -2/q$. At that interval it significantly differs from the second derivative $\Psi''(E)$. Consequently, we see that at least for $q = 4$ we cannot use the normal density as approximation for the central part of the exact spectral density $D(E) = \exp(N \cdot \Psi(E))$

*3.3. Free energy*

In the right column of Fig. 5, we show the graphs of the free energy $f(\beta)$ and its derivatives for the exact spectral density $\Psi(E)$ (solid lines) and the free energy $f_{nv}(\beta)$ corresponding to the spectral density $\Psi_{nv}(E)$ (dashed lines). We see that the graphs of the functions $f(\beta)$ and $f_{nv}(\beta)$ are almost indistinguishable over the whole interval where $\beta$ varies (see the upper right panel in Fig.5). When $\beta < 0.4$, a simple analytical expressions define the function $f_{nv}(\beta)$: $f_{nv}(\beta) = -\ln 2 - \beta^2$. Here the first derivative $f'_{nv}(\beta)$ depends on $\beta$ linearly (see the dashed line on the right middle panel). At $\tilde{\beta}_c \approx 0.39125$ the first derivative $f'_{nv}(\beta)$ has a finite jump and after that it approaches $E_0 = -2$ decreasing smoothly. This jump of the first derivative corresponds to a phase transition of the first kind. Simultaneously, at the same value of $\beta$ the second derivative $f''_{nv}(\beta)$ has an infinite jump. The graphs of the derivatives of the functions $f(\beta)$ and $f_{nv}(\beta)$ differ significantly in the vicinity of the critical inverse temperature $\beta_c \approx 0.44$.

**4. Cubic Ising model**

We do not know the exact expression for the spectral density for the cubic Ising model. That is why we restrict ourselves to analyzing its two approximations. As before by $D_{pl}(E) = \exp(N \cdot \Psi_{pl}(E))$ we denote the spectral density obtained with the aid of the approximation by the polynomials and $D_{nv}(E) = \exp(N \cdot \Psi_{nv}(E))$ is the result of the n-vicinity method. We compare $\Psi_{pl}(E)$ and $\Psi_{nv}(E)$ as well as the corresponding graphs of the free energies and basing on the obtained results make conclusions about qualities of the both approximations.

*4.1. Approximation of exponent $\Psi(E)$ by $\Psi_{pl}(E)$ and $\Psi_{nv}(E)$*

When simulating the spectral density of the cubic Ising model the authors of [7] found that with a good accuracy they could approximate it as

$$D_{pl}(E) \sim \exp(N \cdot \Psi_{pl}(E)), \quad \Psi_{pl}(E) = \ln 2 - a \cdot E^2 + b \cdot E^4 - c \cdot E^6,$$
$$a = 0.1378966,\ b = 0.01459613,\ c = 0.0008365155,\ E \in [-3,+3]. \tag{4.1}$$

They obtained the coefficients by means of the optimal fitting of the simulated exponent to an even polynomial of the six degree. We remind that in the three-dimensional case the energy of the ground state $E_0 = -3$ (see Eq. (1.5).

Let us use the n-vicinity method to calculate the exponent $\Psi_{nv}(E)$ when the number of the nearest neighbors is equal to six ($q = 6$). Solving the minimization problem (3.3) we obtain the following. As long as $E$ is not so large in absolute value, that is when $E < E_0 \approx 0.2113$, the minimum of the function $\varphi(E,x)$ is at the point $x_0 = 1/2$. For such energies, we can calculate $\varphi_m(E)$ using the formula (3.4):

$$\varphi_m(E) = \min_x \varphi(E,x) = -\ln 2 + \frac{E^2}{6}. \tag{4.2}$$

When $E = E_c \approx 0.2113 \cdot E_0$ the point $x_0$ becomes a maximum point of the function $\varphi(E,x)$ and near $x_0$ a minimum appears at a point $x_m(E)$. After that when $E$ moves toward $E_0$ the minimum point $x_m(E)$ moves toward zero, and the minimal value $\varphi_m(E)$ logarithmically tends to zero. After calculation of $\varphi_m(E)$ for all the values of the energy $E \in [-3,0]$ we obtain the required approximation of the exact spectral exponent $\Psi_{nv}(E) = -\varphi_m(E)$ (see. Eq. (3.5)).

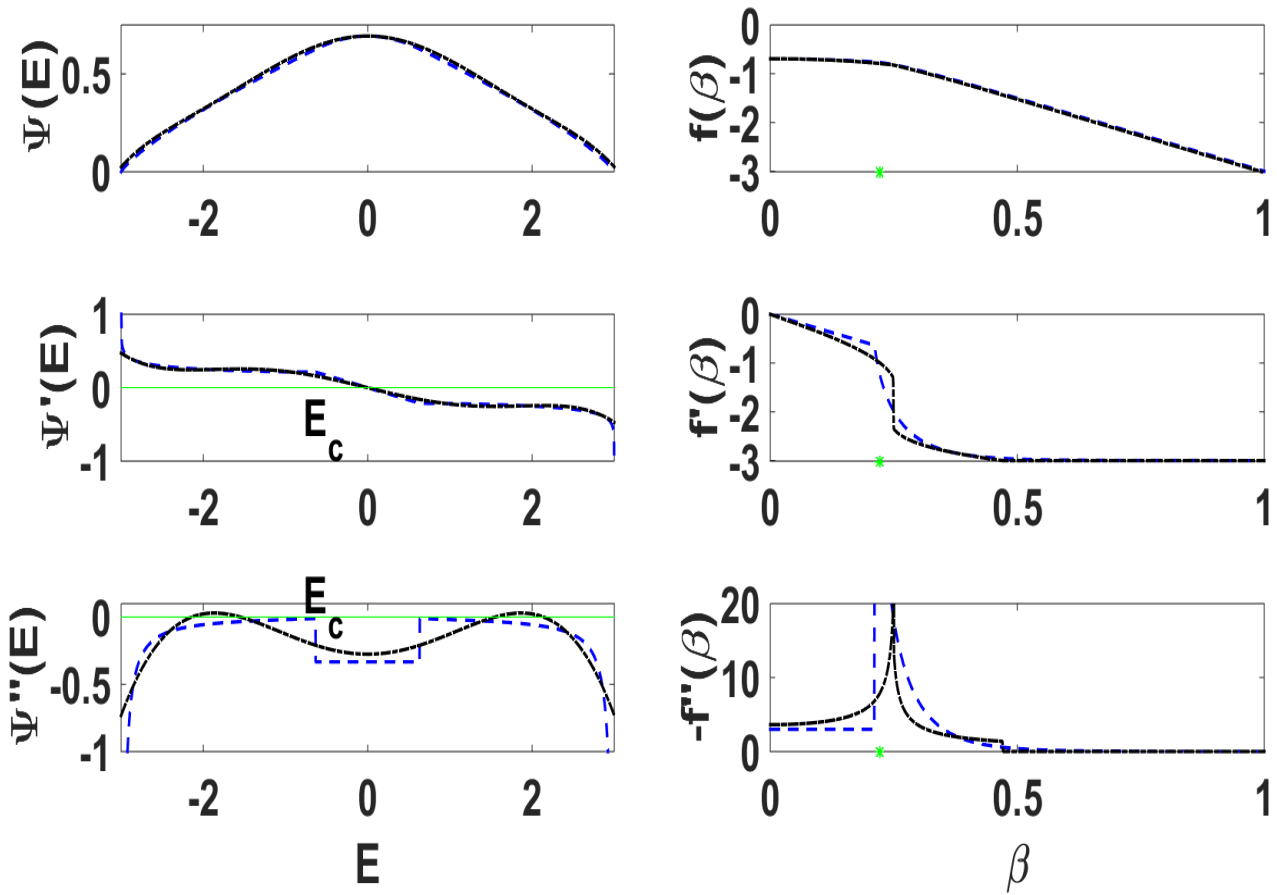

Fig. 6. In left column are $\Psi_{pl}(E)$ (see Eq. (4.1)) and its derivatives (dash-dotted lines) and $\Psi_{nv}(E)$ as well as its derivatives (dashed lines). Right column presents free energies $f_{pl}(\beta)$ (dash-dotted lines) and $f_{nv}(\beta)$ (dashed lines) and their derivative.

In the left column of Fig. 6 we show the graphs of the functions $\Psi_{pl}(E)$ and $\Psi_{nv}(E)$ (the upper panel), their first derivatives $\Psi'_{pl}(E)$ and $\Psi'_{nv}(E)$ (the middle panel) and the second derivatives $\Psi''_{pl}(E)$ and $\Psi''_{nv}(E)$ (the bottom panel). In this figure, the dash-dotted lines correspond to the approximation by polynomials and the dashed lines correspond to the n-vicinity method, respectively.

The graphs of the functions $\Psi_{pl}(E)$ and $\Psi_{nv}(E)$ are very similar, however there are significant differences: We see at $E = E_0$ the polynomial $\Psi_{pl}(E)$ is not equal to zero: $\Psi_{pl}(E_0) > 0$. Consequently, when $N \gg 1$ the spectral density $D_{pl}(E_0)$ can be arbitrary large. In other words, in the framework of the approximation by polynomials the ground state of the system is a many time degenerate state. This result is incorrect. At $E = E_0$ the value of the exponent has to be equal to zero and the function $\Psi_{nv}(E)$ meets this requirement (see Appendix).

In the left middle panel we present graphs of the derivatives $\Psi'_{pl}(E)$ and $\Psi'_{nv}(E)$. In accordance with Eq. (4.2) inside the interval $E_c < E < 0$ the function $\Psi'_{nv}(E)$ decreases linearly. The curve $\Psi'_{nv}(E)$ has a breakpoint at $E_c$ and this leads to a finite jump of the second derivative $\Psi''_{nv}(E)$ (see below). When $E \to E_0$ the derivative $\Psi'_{nv}(E)$ tends to infinity. On the contrary, the derivative $\Psi'_{pl}(E_0)$ is finite when $E \to E_0$. Consequently, in the framework of this approximation the system finds itself in the ground state at a finite temperature (see above). The other problem with $\Psi'_{pl}(E)$ is its non-monotonicity. The derivative $\Psi'_{pl}(E)$ increases inside the interval $-2.1 < E < -1.5$. Consequently, its second derivative is positive and this is forbidden by Eq. (2.4). This means that in the case of the approximation by polynomials the interval $-2.1 < E < -1.5$ is forbidden.

The graphs $\Psi''_{pl}(E)$ and $\Psi''_{nv}(E)$ are in the left bottom panel. The second derivative $\Psi''_{nv}(E)$ tends to negative infinity when $E \to E_0$ and at $E_c$ it has a finite jump that corresponds to a phase transition of the second kind (see below). Inside the interval $[E_c, 0]$ we have $\Psi''_{nv}(E) = -1/3$. On the contrary, the graph of the second derivative $\Psi''_{pl}(E)$ is a smooth curve without singularities. Where the derivative $\Psi''_{pl}(E)$ is positive, a forbidden band of energies appears. A jump over a forbidden band always indicates the phase transition of the first kind. However, nobody predicts such a transition for the cubic Ising model.

*4.2. Free energy*

Let us examine the graphs of the free energies $f_{nv}(\beta)$ and $f_{pl}(\beta)$. In the upper right panel of Fig. 6 we show the graphs of the free energies, in the middle panel there are their first derivatives and the second derivatives are in the bottom panel. The dash-dotted lines we use to present the results of the approximation by polynomials and the dashed lines correspond to the n-vicinity approximation. The marker on the axis of abscissa denotes a critical value of the inverse temperature $\beta_c \approx 0.2216$ obtained for the cubic Ising model in the course of computer simulations [10].

The same as for the planar Ising model the graphs of the functions $f_{nv}(\beta)$ and $f_{pl}(\beta)$ almost do not differ (see the upper right panel of Fig. 6). This is not true for the derivatives of these functions. The function $f'_{nv}(\beta)$ has a breakpoint at $\beta_{nv} \approx 0.2113$ and it has no other singularities. When $\beta \to \infty$ the function $f'_{nv}(\beta)$ tends to the energy of the ground state $E_0 = -3$. Evidently, the second derivative $f''_{nv}(\beta)$ has a finite jump at $\beta = \beta_{nv}$ (see the dashed line in the bottom panel of the figure). In other words, for the cubic Ising model the n-vicinity method predicts the phase transition of the second kind with a finite jump of the heat capacity. The value of the critical inverse temperature $\beta_{nv}$ less than 5% differs from the estimate obtained in the course of computer simulations [10]. As a whole, these results are close to the accepted results for the cubic Ising model.

The other situation is with the approximation of the spectral density by polynomials. In the middle panel, we see a jump of the first derivative $f'_{pl}(\beta)$ at $\beta_{pl} \approx 0.25$ that corresponds to a phase transition of the first kind. After that at $\beta \approx 0.47$ there is another phase transition. Since at that only the second derivative $f''_{pl}(\beta)$ has a small jump (see the dash-dotted line in the bottom panel), this is a phase transition of the second kind. In other words, at a finite temperature the system finds itself in the ground state. We conclude that as a whole the behavior of the function $f_{pl}(\beta)$ and its derivatives contradicts to the physical principles.

## 5. Discussion and conclusions

We analyzed the possibility to calculate the free energy by means of approximations of the spectral density. We performed detailed calculations for the Ising model on the $d$-dimensional hypercube for $d = 2$ (the planar model) and $d = 3$ (the cubic model). Each spin had $q = 2d$ nearest neighbors. We examined two approximations of the exponent of the spectral density $\Psi(E)$ defined in Eq. (2.1). That is we use the n-vicinity method [8, 9] and the same as in [7] we approximated it by polynomials of even degrees.

In the case of the planar model, both approaches give incorrect results: contrary to the classical Onsager solution [3], they predict a phase transition of the first kind. Moreover, when applying the approximation by polynomials the system finds itself in the ground state at a finite temperature and this contradicts to the physical principles (see Subsections 2.3, 2.4).

For the cubic Ising model ($q = 6$) the approximation by polynomials also provides incorrect results. As in the case of the planar model, the system finds itself in the ground state at a finite temperature. In addition, this approximation predicts a phase transition of the first kind. On the contrary, the n-vicinity method much more reasonable. In accordance with the previously obtained results [10], it predicts a phase transition of the second kind at the temperature that is only less than 5% different from the accepted value.

Let us discuss one key question. In Subsection 2.1, we showed that when $E \to E_0$ the exponent $\Psi(E)$ of the exact spectral density tends to zero and its derivative tends to the infinity. When we approximate $\Psi(E)$ by a polynomial $\Psi(E) \approx \Psi_{pl}(E)$ it is easy to fulfill the equality $\Psi_{pl}(E_0) = 0$. However, there are no ways to fulfill the condition $\Psi'_{pl}(E_0) = \infty$. An approximation of the exponent $\Psi(E)$ by a polynomial always leads to a finite derivative $\Psi'_{pl}(E_0) < \infty$. This means that at a finite temperature the system jumps to the ground state (see the end of Subsection 2.3).

The approximation of the exponent $\Psi(E)$ by means of the n-vicinity methods guarantees that when $E \to E_0$ the derivative $\Psi'_{nv}(E) \to \infty$ (see the dashed lines in the left middle panels of Figs. 5 and 6). In Appendix we show that in general case the second requirement $\lim_{E \to E_0} \Psi_{nv}(E) = 0$ is also fulfilled when $q > 4\ln 2$. However when $q < 16/3$ the n-vicinity method works poor (see Section 3 and [9]). We hope to expand the limits of the applicability of this method replacing the Gaussian distribution by a more general approximation where not only the first two moments but also higher moments are taken into account. We think that is possible since even a small correction of the Gaussian approximation by the Edgeworth expansion [11] that includes the third moment leads to positive results. For the high-dimensional Ising models ($3 \leq d \leq 7$) the Edgeworth expansion improves the agreement with the results of computer simulations by 20%. The publication of these results follows.

Summing the aforesaid, we conclude that the approximation of the spectral density by polynomials is not suitable when calculating the free energy. As shown in Appendix, when $q < 4\ln 2$ one cannot apply the current version of the n-vicinity method. However, when $q$ is large, that is $q > 16/3$, it provides the results that are in good agreement with the widely recognized results. We hope to refine this method in order to extend the boundaries of its applicability.

**Acknowledgments**

The work was supported by the Russian Basic Research Foundation (grants 16-01-00626 and 18-07-00750). The authors are grateful to Dr. Inna Kaganova for help during the course of the work.

**Appendix**

Let $\Psi_{nv}(E)$ be a result the n-vicinity approximation of the exponent of the exact spectral density $\Psi(E)$ for the Ising model with $q$ nearest neighbors. The ground state of the Ising model is doubly degenerate and consequently $D_{nv}(E_0) = \exp(N \cdot \Psi_{nv}(E_0)) = 2$. It is evident that when $N \to \infty$ we can fulfill the equality $D_{nv}(E_0) = 2$ only if $\Psi_{nv}(E) \to 0$ when $E \to E_0$. Let us find out when the last equality is valid.

We need to minimize the function $\varphi(E, x)$ (see Eq. (3.3)):

$$\varphi_m(E) = \min_{0 \leq x \leq 1} \varphi(E, x), \text{ where } \varphi(E, x) = S(x) + \frac{q}{4}\left(1 - \frac{1-\varepsilon}{4x(1-x)}\right)^2, \quad \varepsilon = E/E_0, \tag{A1}$$

and at that we need $\varphi_m(E) \to 0$ when $E \to E_0$.

Equating to zero the first derivative of the function $\varphi(E, x)$ with respect to $x$ we obtain the following equation for the coordinate of the minimum point:

$$z(x)[1 - z(x)] = \alpha(x), \text{ where } \alpha(x) = \frac{2x(1-x)}{q(1-2x)} \ln \frac{1-x}{x} \text{ and } z(x) = 1 - \frac{1-\varepsilon}{4x(1-x)}. \tag{A2}$$

On the interval $[0, 1/2]$ the function $\alpha(x)$ increases monotonically from zero to $\alpha(1/2) = 1/q$. When $q > 1$ at any $x$ we have $\alpha(x) < 1$. Consequently, we can use $\alpha(x)$ as a small parameter for a series expansion. Then when solving Eq. (A2) with respect to $z(x)$ we obtain a new equation

$$z_1(x) = \frac{1-\sqrt{1-4\cdot\alpha(x)}}{2} \approx \alpha(x).$$

This equation relates the value of $\varepsilon$ with a coordinate $x_\varepsilon$ where the function $\varphi(E,x)$ has its minimum. When $\varepsilon \to 1$ the point $x_\varepsilon$ tends to zero: $x_\varepsilon \to 0$. Consequently, when $\varepsilon \to 1$ the solution of the problem (A1) tends to zero:

$$\lim_{\varepsilon\to 1}\varphi_m(\varepsilon) \approx \lim_{\varepsilon\to 1}\left(x_\varepsilon \cdot \ln x_\varepsilon + (1-x_\varepsilon)\ln(1-x) + \frac{q}{4}\alpha(x_\varepsilon)\right) = 0.$$

On the other hand, always there is a trivial minimum at the point $x_0 = 1/2$. At $\varepsilon \to 1$ its value is

$$\lim_{\varepsilon\to 1}\varphi(E,x_0) = \lim_{\varepsilon\to 1}\left(S(x_0) + \frac{q}{4}\varepsilon^2\right) = -\ln 2 + \frac{q}{4}.$$

In order that $\Psi_{nv}(E)$ tends to zero when $E \to E_0$ the limit $\lim_{\varepsilon\to 1}\varphi_m(\varepsilon)$ necessarily has to be less than $\lim_{\varepsilon\to 1}\varphi(E,x_0)$. From this we obtain

$$q > 4\ln 2.$$

This inequality defines the boundary for the parameter $q$. When $q < 4\ln 2$, we cannot use the n-vicinity method.